\documentclass[prl,twocolumn,amsmath,amssymb,superscriptaddress]{revtex4}% Physical Review B

\usepackage{graphicx}% Include figure files
\usepackage{dcolumn}% Align table columns on decimal point
\usepackage{bm}% bold math
\begin{document}

%\preprint{}

\title{First direct measurement of the intrinsic dipole moment in pear-shaped thorium isotopes}
% Force line breaks with \\

\author{M.~M.~R.~Chishti}
\affiliation{SUPA, School of Computing, Engineering and Physical Sciences, University of the West of Scotland, Paisley, PA1 2BE, United Kingdom.}

\author{D.~O'Donnell}
%\thanks{}
\email{david.odonnell@uws.ac.uk}
\affiliation{SUPA, School of Computing, Engineering and Physical Sciences, University of the West of Scotland, Paisley, PA1 2BE, United Kingdom.}

\author{G.~Battaglia}
\affiliation{SUPA, Department of Physics, University of Strathclyde, Glasgow, G4 0NG, United Kingdom.}

\author{M.~Bowry}
\affiliation{SUPA, School of Computing, Engineering and Physical Sciences, University of the West of Scotland, Paisley, PA1 2BE, United Kingdom.}

\author{D.~A.~Jaroszynski}
\affiliation{SUPA, Department of Physics, University of Strathclyde, Glasgow, G4 0NG, United Kingdom.}

\author{B.~S.~Nara Singh}
\affiliation{SUPA, School of Computing, Engineering and Physical Sciences, University of the West of Scotland, Paisley, PA1 2BE, United Kingdom.}

\author{M.~Scheck}
\affiliation{SUPA, School of Computing, Engineering and Physical Sciences, University of the West of Scotland, Paisley, PA1 2BE, United Kingdom.}

\author{P.~Spagnoletti}
\affiliation{SUPA, School of Computing, Engineering and Physical Sciences, University of the West of Scotland, Paisley, PA1 2BE, United Kingdom.}

\author{J.~F.~Smith}
\affiliation{SUPA, School of Computing, Engineering and Physical Sciences, University of the West of Scotland, Paisley, PA1 2BE, United Kingdom.}

\date{\today}% It is always \today, today,

\begin{abstract}
It is now well established that atomic nuclei composed of certain combinations of protons and neutrons can adopt reflection-asymmetric, or octupole-deformed, shapes at low excitation energy. These nuclei show promise in the search for a permanent atomic electric dipole moment, the existence of which has implications for physics beyond the Standard Model. Theoretical studies have suggested that certain isotopes of thorium may have the largest octupole deformation. However, due to experimental challenges, the extent of the octupole collectivity in the low-energy states in these thorium nuclei has not yet been demonstrated. This paper reports measurements of the lifetimes of low-energy states in $^{228}$Th ($Z = 90$) undertaken using the mirror symmetric centroid difference method, which is a direct electronic fast-timing technique. Lifetime measurements of the low-lying $J^\pi = 1^-$ and $3^-$ states, which are the first for a thorium isotope, have allowed the $B(E1)$ rates and the intrinsic dipole moment to be determined. The results are in agreement with those of previous theoretical calculations allowing the extent of the octupole deformation of $^{228}$Th to be estimated. This study indicates that the nuclei $^{229}$Th and $^{229}$Pa ($Z = 91$) may be good candidates for the search for a permanent atomic electric dipole moment.

\end{abstract}

\maketitle
%\section{Introduction}
Bohr and Mottelson interpreted observed spectra of deformed atomic nuclei as the manifestation of rotational and vibrational degrees of freedom~\cite{Bohr,BohrMottelson}. In this pioneering work, the excitations were understood as rotational or vibrational modes of quadrupole deformed axially-symmetric nuclei. Since then it has became clear that some nuclei may also adopt reflection asymmetric shapes, such as those resulting from long-range octupole-octupole residual interactions. The operator associated with such interactions has negative parity, which leads to the nucleus adopting shapes where reflection symmetry is broken. The distinctive shape associated with static octupole deformation has led to such nuclei being referred to as pear shaped, where the nuclear density is higher at one pole than the other. Such octupole deformation can result in a separation of the centres of charge and mass which can result in a sizeable electric dipole ($E1$) moment, which will be observable through enhanced $E1$ transitions between nuclear states~\cite{Butler_dipole_moments}.  

Reflection asymmetric nuclei have attracted attention in recent years in the search for a permanent atomic electric dipole moment (EDM). The Standard Model predicts a vanishingly small, but non-zero, EDM. To date, no evidence has been found that supports this prediction, despite several experiments having been undertaken using different methods~\cite{EDM_2002,EDM_2011,EDM_2014,EDM_2017,EDM_2018}. This is because the EDM predicted by the Standard Model is far below the sensitivity of current experimental capabilities. Nevertheless, by placing upper limits on an EDM, these studies have successfully constrained any proposed extensions of the Standard Model predicting a significantly larger EDM value. The ACME II experiment, which has recently improved on the upper limit on the EDM value by an order of magnitude~\cite{EDM_2018}, involves laser spectroscopy of thorium monoxide molecules to measure the EDM. Some isotopes of thorium are predicted to exhibit an enhanced octupole collectivity at low excitation energies~\cite{Agbemava}. The use of such nuclei in the search for an EDM would be particularly attractive because the induced $E1$ moment resulting from the reflection asymmetric shape enhances the nuclear Schiff moment~\cite{Schiff}, and therefore the atomic EDM, by two to three orders of magnitude~\cite{Spevak_dipole,Ellis_dipole}. In particular, Ref.~\cite{Spevak_dipole} shows that the Schiff moment in the laboratory frame exhibits a quadratic dependence on the extent of the octupole deformation. 

One nucleus identified early as an ideal candidate with which to search for an EDM is $^{229}$Pa ($Z = 91$). Possessing an odd number of protons, this nucleus was predicted to have a low-lying parity doublet in which two states with the same total angular momentum are found almost degenerate in energy~\cite{Chasman}. Such a doublet, which is regarded as a signature of octupole correlations, was first reported by Ahmad {\it et al.}~\cite{Ahmad1982} in 1982. More recently, however, doubt has been cast upon the existence of this doublet due to a lack of conclusive evidence~\cite{Grafen,Losch,Ahmad2015}. Should $^{229}$Pa indeed be octupole deformed in the ground state it is expected that the even-even nucleus $^{228}$Th ($Z = 90$), which constitutes the core of $^{229}$Pa to which an unpaired proton is coupled, would also exhibit characteristics consistent with enhanced octupole correlations. Another nucleus where an enhanced nuclear Schiff moment is predicted in thorium-229, which corresponds to an unpaired neutron coupled to a $^{228}$Th core~\cite{Flambaum_229Th_EDM}. This further emphasises the importance of determining the octupole deformation in $^{228}$Th.

A signature of reflection asymmetry in even-even nuclei is the presence of low-lying negative parity states. Such states, with total angular momentum values corresponding to $1~\hbar$ and $3~\hbar$, were first identified in $\alpha$-decay spectroscopy measurements in the 1950s~\cite{Stephens1,Stephens2}. Electromagnetic transitions between these negative parity states and members of the ground state band are also good indications of octupole collectivity. For instance, the observation of enhanced electric octupole ($E3$) transitions between the low-lying $J = 3^-$ state and the $J = 0^+$ ground state is considered as an unambiguous signature of octupole collectivity~\cite{Robledo_Butler}. To date only two known nuclei have been found to exhibit both signatures: $^{224}$Ra~\cite{Gaffney} and $^{226}$Ra~\cite{Wollersheim_octupole}. Nevertheless, there are a several actinide nuclei that have low-lying negative parity states; theoretical studies have suggested they should have appreciable octupole deformations in the ground state~\cite{Egido_Robledo1989,Agbemava}. Indeed,  density functional theories used in Ref.~\cite{Agbemava} suggest that thorium nuclei are likely to have the largest octupole deformations of all the actinides. However, disagreement exist between different theoretical models on which thorium nucleus should exhibit the largest octupole collectivity. In the work of Agbemava {\it et al.}~\cite{Agbemava}, pronounced quadrupole and octupole deformations are expected in the ground states of $^{226}$Th and $^{228}$Th and the octupole deformation parameter $\beta_3$ is maximised for the latter. In this paper we investigate the extent of octupole collectivity in $^{228}$Th by measuring the enhancement of electric dipole transitions from the low-lying $J = 1^-$ and $3^-$ states. 

The observation of enhanced $E1$ transitions connecting negative and positive parity states is generally considered a good indication of reflection asymmetry, albeit with some ambiguity~\cite{Butler_dipole_moments,Butler_review1996}. One source of ambiguity arises from the enhancement of electric dipole transitions that are not unique to octupole deformed nuclei but also manifest in well-deformed reflection-symmetric nuclei due to the presence of low-lying octupole vibrational states~\cite{Butler_review1996}. This is further complicated because of the microscopic shell contribution to the dipole moment in the intrinsic frame effectively negates the macroscopic contribution from the reflection asymmetric shape of the nucleus, which was first identified by Butler and Nazarewicz~\cite{Butler_dipole_moments}. The most striking example of this is observed in the work of Gaffney {\it et al.}~\cite{Gaffney} where they reported measurements of a large $E3$ strength in $^{224}$Ra. This indicates a sizeable octupole collectivity in the ground state, yet a small $E1$ strength, consistent with reflection-symmetric nuclei, was also reported.

In our study we measure the lifetimes of low-lying excited states of $^{228}$Th using electronic fast timing techniques based on fast scintillating $\gamma$-ray detectors. This represents the first measurement of the absolute electric dipole transition rates in any thorium isotopes, for the low-lying negative-parity states characteristic of octupole collectivity. When compared with calculations available in the literature~\cite{Robledo_Butler} we conclude that the low-energy structure of $^{228}$Th is consistent with a strongly quadrupole-deformed nucleus with an octupole deformation comparable with that reported in $^{224}$Ra~\cite{Gaffney}. Our findings suggest that the odd-$A$ nuclei $^{229}$Th and $^{229}$Pa may be good candidates with which to search for a permanent EDM.

\begin{figure}
\centering
\includegraphics[width=0.35\textwidth,angle=270,clip]{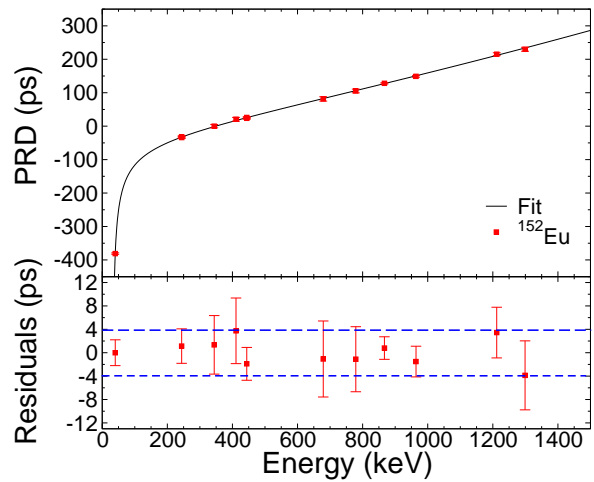}
\caption{Top shows the $\gamma-\gamma$ time-walk dependence with $\gamma$-ray energy of the fast-timing apparatus. Data points correspond to transitions depopulating prompt states in $^{152}$Sm and $^{152}$Gd following the $\beta$-decay of $^{152}$Eu. Error bars are smaller than the symbols for the data points. Bottom shows the differences between the data and the fitted prompt response function with the dashed line indicating two standard deviations.}
\label{PRD}
\end{figure}

%\section{Experimental results}
\section*{Results}
\subsection*{Measurement of the lifetimes of excited states of $^{228}$Th}
\begin{figure}
\centering
\includegraphics[width=0.28\textwidth,angle=270,clip]{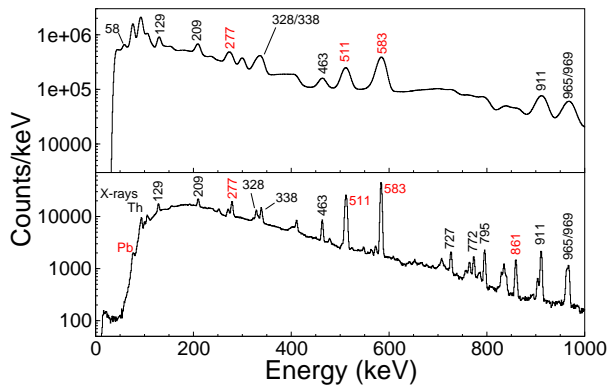}
\caption{Gamma-ray spectra resulting from the decay of $^{232}$Th. The top spectrum shows $\gamma$ radiation detected with the Start LaBr detector in coincidence with radiation detected in the Stop LaBr detector. The bottom spectrum shows the $\gamma$-ray spectrum detected with a high-purity germanium detector in coincidence with $\gamma$ rays detected in both the Start and Stop detectors. Peaks labelled in black and red are associated with the decay of excited states in $^{228}$Th and $^{208}$Pb, respectively.}
\label{gammas}
\end{figure}

Spectra of $\gamma$ radiation detected with the Start LaBr and HPGe detectors are shown in Figure~\ref{gammas}. The definition of the detectors can be found in the Methods section of this paper. Since data are only recorded when a TAC pulse is generated, the LaBr spectrum of Figure~\ref{gammas} represents the projection of a two-dimensional $\gamma-\gamma$ (or Start-Stop) coincidence matrix. The HPGe spectrum, however, is the projection of a three-dimensional $\gamma-\gamma-\gamma$ (Start-Stop-HPGe) coincidence matrix. It is clear from the spectrum that the majority of the $\gamma$ radiation observed is the result of decays of excited states in $^{208}$Pb ($Z = 82$) and $^{228}$Th. The former is the result of the $\beta$-decay of $^{208}$Tl ($Z = 81$) as the final stage of the $^{232}$Th decay chain, while excited states of $^{228}$Th are populated as the result of $\beta$-decay of $^{228}$Ac ($Z = 89$). There are a few intense $\gamma$ rays associated with the $^{232}$Th decay chain that are not observed in the spectra of Fig.~\ref{gammas} due to the decays having unity multiplicity and, therefore, do not generate a TAC output. 

\begin{figure}
\centering
\includegraphics[width=0.2\textwidth,angle=270,clip]{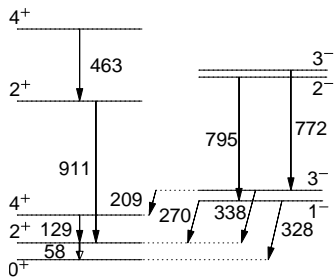}
\caption{A partial level scheme showing the excited states and $\gamma$-ray transitions of relevance to this study. The $\gamma$-ray transitions are indicated by vertical arrows with the corresponding energy in units of keV shown.}
\label{level_scheme}
\end{figure}

\begin{figure}
\centering
\includegraphics[width=0.38\textwidth,angle=270,clip]{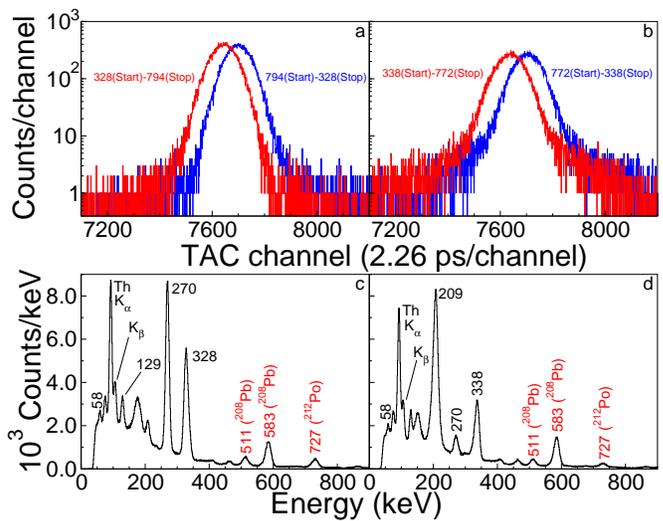}
\caption{$\mathbf{a}$ TAC spectra showing delayed and anti-delayed coincidences between 795~keV and 328~keV transitions populating and depopulating the $J^\pi = 1^-$ state; $\mathbf{b}$ similar to $\mathbf{a}$ but for the 772~keV and 338~keV transitions populating and depopulating the $J^\pi = 3^-$ state; $\mathbf{c}$ spectrum of $\gamma$ rays in coincidence with 795~keV transition feeding the $J^\pi = 1^-$ state; $\mathbf{d}$ $\gamma$ rays in coincidence with 772~keV transitions feeding the $J^\pi = 3^-$ state.}
\label{TAC}
\end{figure}

The excited states and $\gamma$-ray transitions of interest in $^{228}$Th are shown in Figure~\ref{level_scheme}. To measure the lifetime of the $J^\pi = 1^-$ state, the time difference between the arrival of 795~keV and 328~keV transition photons (as they populate and depopulate the state, respectively) is measured with the TAC. The corresponding TAC spectra are shown in Figure~\ref{TAC}a while a sample of the $\gamma$ rays observed in coincidence with the 795~keV transition is presented in Figure~\ref{TAC}c. Peaks corresponding to the 270~keV and 328~keV transitions de-exciting the $1^-$ state clearly dominate the spectrum of Fig.~\ref{TAC}c. In addition, some background peaks corresponding to other intense transitions in the $^{232}$Th decay chain can be observed. As a result, the peaks of interest for lifetime measurements sit  on top of a background dominated by events corresponding to Compton scattered background $\gamma$ rays. The method outlined by Ansari {\it et al.}~\cite{Ansari} has been employed to correct for the effect of this background. Following correction, the lifetime of the $J^\pi = 1^-$ state is measured to be $\tau = 3.3(21)$~ps.   

Figure~\ref{TAC}d shows $\gamma$ radiation observed in coincidence with 772~keV transitions which feed the $J^\pi = 3^-$ state. The 209~keV and 338~keV transitions, which have been established to deexcite the $3^-$ state, dominate this spectrum in addition to some background peaks. The 209~keV peak is significantly wider than that at 338~keV (and indeed the 270~keV peak in Fig.~\ref{TAC}c) suggesting the presence of additional background. It is likely that this is due to backscattered photons originating from the other nearby detectors. The TAC spectra corresponding to the 772~keV feeder and the 338~keV decay transitions are shown in Fig.~\ref{TAC}b. The background-corrected lifetime for the $3^-$ state is found to be $\tau = 13.2(23)$~ps. In addition to the lifetimes of the low-lying negative parity states, lifetimes of three positive-parity states have been measured and listed in Table~\ref{Table}.  
\begin{table}
\caption{Properties of levels and transitions in $^{228}$Th as measured in this work.}
\begin{ruledtabular}
\begin{tabular}{ccccccc}
$E$& $J^\pi_i \rightarrow J^\pi_f$ & $\tau$ & $B(E1)$ & $\vert D_0 \rvert$ & $B(E2)$ & $Q_0$ \\
{[keV]} & & {[ps]} & {[$10^{-3}e^2$fm$^2$]} & {[$e$fm]} & {[$10^4e^2$fm$^4$]} & {[$e$fm$^2$]}\\
\hline
58 & $2^+ \rightarrow 0^+$& 587(12) & & & 1.40(3) & 853(17)\\
187 & $4^+ \rightarrow 2^+$ & 260(14) & & & 2.40(13) & 924(51)\\
328 & $1^- \rightarrow 0^+$ & 3.3(21) & 2.4(15) & 0.17(8) & & \\
396 & $3^- \rightarrow 2^+$ & 13.2(23) & 0.9(2) & 0.10(2) & & \\
969 & $2^+ \rightarrow 2^+$ & 3.8(21) & & & 0.021(12) & 55(27)\\
969 & $2^+ \rightarrow 0^+$ & 3.8(21) & & & 0.009(5) & 44(22)\\
\end{tabular}
\label{Table}
\end{ruledtabular}
\end{table}

The lifetimes of the first $J^\pi = 2^+$ and $4^+$ states have previously been measured using the delayed-coincidence technique, where the difference in time between detection of a populating $\alpha$ particle and the de-exciting $\gamma$-ray was measured~\cite{Ton_alpha_gamma}. Our results are in good agreement with these studies confirming a large quadrupole deformation in the ground state. We have also measured the lifetime of the $J^\pi = 2^+$ state for an excitation energy of 969~keV. Previous studies of the high-spin states of $^{228}$Th have identified the state as the bandhead of a $K^\pi = 2^+$ $\gamma$-vibrational band~\cite{Weber_gamma_band}. The reduced transition probability, $B(E2; 2_\gamma^+ \rightarrow 0^+) = 1.1(6)$~Wu, is similar to the $B(E2; 2_\gamma^+ \rightarrow 0^+) = 1.3(5)$~Wu value measured by Gaffney {\it et al.}~\cite{Gaffney} for the $\gamma$-vibrational bandhead in $^{224}$Ra. In addition, the $\frac{B(E2; 2_\gamma^+ \rightarrow 0^+)}{B(E2; 2_\gamma^+ \rightarrow 2^+)}$ ratio is consistent, within uncertainties, with the limit of 0.7 expected from the Alaga rules~\cite{Alaga}, which further supports the assignment of this level as the $\gamma$-vibrational bandhead in $^{228}$Th.

\section*{Discussion}
The measured lifetime of the $J^\pi = 1^-$ state allows for a $B(E1;1^- \rightarrow 0^+) = 1.0(6) \times 10^{-3}$~Wu value to be extracted. A $B(E1)$ rate of this magnitude is consistent with enhanced octupole collectivity, but is similar to the $B(E1;1^- \rightarrow 0^+) < 1.5 \times 10^{-3}$~Wu value reported by Gaffney {\it et al.}~\cite{Gaffney} for the octupole-vibrating nucleus $^{220}_{86}$Rn. Few theoretical predictions exist for the $E1$ strength of low-energy states of the thorium nucleus. A comprehensive study of the dipole strength in actinide nuclei, performed by Butler and Nazarewicz~\cite{Butler_dipole_moments}, predicts the $E1$ strength  only for the high-spin states of $^{228}$Th. A more recent study, performed by Robledo and Butler~\cite{Robledo_Butler}, considered the coupling of the quadrupole and octupole collective degrees of freedom to predict $B(E1;1^- \rightarrow 0^+) = 1.2 \times 10^{-3}$~Wu for $^{228}$Th. 

The intrinsic dipole moment, $D_0$, can be calculated directly from the measured $B(E1)$ rates according to the rotational model formula:
\begin{equation}
B(E1;J_i \rightarrow J_f) = \frac{ 3 }{4 \pi} {D_0}^2 {\langle J_i K_i 10 | J_f K_f \rangle}^2,
\end{equation} where $\langle J_i K_i 10 | J_f K_f \rangle$ is the Clebsch-Gordan coefficient. The use of a rotational model formula is appropriate for a nucleus such as $^{228}$Th, which exhibits clear signs of rotational motion ($Q_0 = 853(17)~e$fm$^2$ and $E(4_1^+)/E(2_1^+) = 3.22$). This formula is commonly applied to the actinide region, regardless of how rotational the nuclei appear, to obtain a consistent measure of the dipole moment from available data. Our measurement results in a value of $D_0 = 0.17(8)~e$fm for the $J^\pi = 1^-$ state. This $D_0$ value, while consistent within one standard deviation, is larger than the values reported for any of the higher spin states of $^{228}$Th  extracted indirectly from observed $B(E1)/B(E2)$ ratios~\cite{SchulerTh,Ackermann_226Th}. This dipole moment is considerably larger than the prediction of $D_0 = 0.07~e$fm for $J^\pi = 8^-$ by Butler and Nazarewicz~\cite{Butler_dipole_moments} indicating that the dipole moment, and possibly, the octupole collectivity in this nucleus is underestimated in their calculations. 

The intrinsic dipole moment of $D_0 = 0.10(2)~e$fm calculated in the current work for the $J^\pi = 3^-$ state in $^{228}$Th reproduces the results of Ackermann {\it et al.}~\cite{Ackermann_226Th} in which a lower limit $(D_0 \geq 0.1~e$fm$)$ is determined by means of $B(E1)/B(E2)$ ratios. The lower limit in the previous study corresponds to a maximum value of the observed branching of the $3^- \rightarrow 1^-$ $E2$ transition. It is worth noting that the value of $D_0$ measured for the $J^\pi = 3^-$ state in our work is in good agreement with the values established for higher-spin states in the octupole band of $^{228}$Th which range from $0.11-0.13$~\cite{Ackermann_226Th,SchulerTh}.

The fact that the calculations of Robledo and Butler~\cite{Robledo_Butler} reproduce the experimentally-determined $B(E1;1^- \rightarrow 0^+)$ value suggests that they can be used to estimate the extent of octupole collectivity in the low-energy states of $^{228}$Th. In the case of the ground state of $^{228}$Th, the calculations of Ref.~\cite{Robledo_Butler} predict a pronounced minimum in the potential energy corresponding to an octupole moment $Q_3 = 3500~e$fm$^3$. These calculations overestimate the octupole moment in $^{224}$Ra, which was measured to be $Q_3 = 2520(90)~e$fm$^3$~\cite{Gaffney}, by a factor of $\sim15\%$. Scaling the octupole moment for $^{228}$Th to $2950~e$fm$^3$ and using the measured quadrupole moment $Q_0(2^+) = 853(17)~e$fm$^2$ and the equations described by Leander and Chen~\cite{LeanderChen} allows for the calculation of the quadrupole and octupole deformation parameters $\beta_2 = 0.19(1)$ and $\beta_3 = 0.11(2)$. These values are within two standard deviations of the theoretical values of $\beta_2 = 0.21$ and $\beta_3 = 0.15$ reported by Agbemava {\it et al.}~\cite{Agbemava} for $^{228}$Th. Comparing the values obtained in this work with those reported for $^{224}$Ra~\cite{Gaffney} ($\beta_2 = 0.154$ and $\beta_3 = 0.097$) indicates that odd-$A$ nuclei composed of a $^{228}$Th core such as $^{229}$Pa and $^{229}$Th may be superior candidates for the search for a nuclear EDM. The increased $\beta_3$ value for $^{228}$Th means that the collective Schiff moment may be larger in $^{229}$Pa and $^{229}$Th compared with odd-$A$ Rn and Ra nuclei currently attracting attention in searches for an atomic EDM~\cite{Parker_Ra225_EDM,Bishof_Ra225_EDM}.

In summary, the electronic fast-timing technique known as mirror symmetric centroid difference method has been used to directly measure the lifetimes of several low-energy states of the nucleus $^{228}$Th following  population via  $\beta$-decay of the ground state of $^{228}$Ac. The measured states include the first and second members, $J^\pi = 1^-$ and $3^-$, of the $K^\pi = 0^-$ band associated with octupole correlations in this nucleus.  This is the first time the lifetimes of these states have been probed in thorium isotopes. The measured $B(E1)$ rates are consistent with the presence of enhanced octupole correlations at low energy, and with $^{228}$Th having large quadrupole collectivity. Indeed, the estimated $\beta_3$ value for $^{228}$Th indicates that an enhanced Schiff moment may be observed in $^{229}$Th and $^{229}$Ac indicating these nuclei may be superior to Rn and Ra nuclei currently employed in the search for an atomic EDM. 

%\begin{figure}
%\centering
%\includegraphics[width=0.4\textwidth,angle=270,clip]{218Th_paper_level_scheme.eps}
%\caption{The level scheme representing the decay of excited states of $^{218}$Th established in the present work. The non-tentative spin and parity assignments reported in Ref.~\cite{Bonin} have been included. Transition and level energies are given in units of keV and the widths of the arrows are proportional to the measured intensities of the transitions. The unfilled region of the arrows is indicative of internal conversion.}
%\label{level_scheme}
%\end{figure}

%\section{Discussion}

\section*{Methods}
\subsection*{Experimental details and detectors}
%\section{Experimental details}
Excited states of $^{228}$Th are populated by the $\beta$-decay of the ground state of $^{228}_{89}$Ac following the $\alpha$-decay of the $^{232}$Th nucleus. A foil of natural $99.5\%$ purity thorium with dimensions of $50 \times 50 \times 0.05$~mm acted as a source of $^{232}$Th nuclei with an activity corresponding to approximately 6~kBq. Due to being in a state of secular equilibrium the foil provided  $\sim$6~kBq of $^{228}$Ac. The thorium foil was located equidistant between two cerium-doped lanthanum bromide (LaBr) detectors, the front faces of which were separated by 10~mm. The cylindrical LaBr crystals each had a diameter and length of 25.4~mm and are coupled to Hamamatsu R9420 photomultiplier tubes. The detectors were located 60~mm from the thorium foil. An Ortec GEM P-type high-purity germanium (HPGe) detector was orientated $90^\circ$ relative to the LaBr-LaBr detector axis. This detector had a relative efficiency of 59\% and an endcap diameter of 70~mm. Signals from the LaBr detectors were optimised for timing measurements and connected to a time-to-amplitude converter (TAC), which produced an output voltage proportional to the difference in time between detection of two detected $\gamma$-ray photons, respectively. 

Waveform traces from each of the detectors and the TAC were recorded using a CAEN V1725 digitiser and stored using the MIDAS data acquisition software. Custom software was used to reduce the data by extracting the pulse heights from the traces using a moving window deconvolution algorithm. The reduced data were subsequently analysed using the Root data analysis framework~\cite{Root}.

The lifetimes of excited states were measured according to the mirror symmetric centroid difference (MSCD) method outlined by Regis {\it et al.}~\cite{Regis_MSCD}. This method involves two $\gamma$-ray detectors, the signals of which are processed by a constant-fraction discriminator (CFD). The resultant logic signal from one CFD is used as the Start input of the TAC while the other is artificially delayed before providing the Stop signal for the TAC. The method involves using the Start detector to detect the $\gamma$-ray photons populating an excited state of interest and the Stop detector to detect $\gamma$ rays depopulating the state. As a result, a spectrum of TAC values with a Gaussian distribution is obtained when the lifetime of the state is sufficiently short ($\tau \leq 150$~ps). For longer lifetimes the Gaussian distribution has an exponential tail. This configuration is known as the delayed configuration. In the anti-delayed configuration the Start detector is used to detect the depopulating transition while the Stop detects the populating transition. The difference between the first moments of the delayed and anti-delayed TAC spectra ($\Delta C$) is proportional to the lifetime of the state. However, a natural `time-walk' is associated with the CFD, which must be accounted for to obtain accurate lifetimes. The effect of the time-walk is such that $\Delta C$ values will increase as the energy of the detected transitions increases regardless of the lifetime of the state. The time-walk can be quantified by measuring $\Delta C$ as a function of energy for precisely known picosecond transitions. There are a number of suitable states in the $^{152}$Sm and $^{152}$Gd nuclei that are populated as a result of the $\beta$-decay of a standard $^{152}$Eu $\gamma$-ray calibration source. The resulting $\Delta C$ values obtained using the prompt states in $^{152}$Sm and $^{152}$Gd are shown in Figure~\ref{PRD} as a function of the energy of the depopulating transition. The resulting curve fitted to the $^{152}$Eu data is known as the prompt response function (typically referred to as PRD) of the fast-timing apparatus, which is used to correct for the CFD time-walk. The lifetime is calculated as
$\tau = (\Delta C - \Delta PRD)/2$, 
where $\Delta PRD$ is the difference between PRD values for the $\gamma$-ray energies detected with the Start and Stop detectors, respectively.

\bibliography{daresbury_bibliography}
%\bibliography{MyCollection}

\section*{Acknowledgments}
The authors would like to acknowledge useful discussions with Luis Robledo who was the recipient of a SUPA Distinguished Visitor grant awarded to UWS. Financial support for this work has been provided by the Scottish Funding Council (SFC) and the UK Science and Technology Facilities Council (STFC). We acknowledge support of the U.K. EPSRC (Grant No. EP/J018171/1, EP/J500094/1 and EP/N028694/1), the EC’s LASERLAB-EUROPE (Grant No. 654148), EuCARD-2 (Grant No. 312453), EuPRAXIA (Grant No. 653782), ARIES (Grant No. 730871) and the Extreme Light Infrastructure (ELI) European Project.

\section*{Author Contributions}
D.O'D. conceived the investigation; M.M.C., D.O'D. and G.B. set up the instrumentation; M.M.C., D.O'D., G.B. and P.S. performed the data analysis; M.M.C., D.O'D. and M.S. interpreted the results; D.O'D., M.B., D.A.J., B.S.N.S., M.S., P.S. and J.F.S. contributed to writing the manuscript.

\section*{Additional Information}
{\bf Competing Interests:} The authors declare no competing interests.

\end{document}